\documentclass[aoas,MSNbibl,nameyear,seceqn,dvips]{arximspdf}
\usepackage{graphicx}

\doi{10.1214/11-AOAS509}
\volume{6}
\issue{1}
\pubyear{2012}
\firstpage{210}
\lastpage{228}

\makeatletter

\newcommand{\bbr}{\mathbb{R}}

\newproclaim{Result}{Result}
\newproclaim{Remark}{Remark}

\makeatother

\begin{document}
\begin{frontmatter}

\title{Properties of design-based estimation under stratified spatial
sampling with application to canopy coverage estimation}
\runtitle{Design-based estimation under stratified spatial sampling}

\begin{aug}
\author[A]{\fnms{Lucio} \snm{Barabesi}\ead[label=e1]{barabesi@unisi.it}},
\author[A]{\fnms{Sara} \snm{Franceschi}\corref{}\ead[label=e2]{franceschi2@unisi.it}}
\and
\author[A]{\fnms{Marzia} \snm{Marcheselli}\ead[label=e3]{marcheselli@unisi.it}}
\runauthor{L. Barabesi, S. Franceschi and M. Marcheselli}
\affiliation{University of Siena}
\address[A]{Economics and Statistics Department\\
University of Siena\\
Piazza San Francesco 17\\
53100, Siena\\
Italy\\
\printead{e1}\\
\hphantom{E-mail: }\printead*{e2}\\
\hphantom{E-mail: }\printead*{e3}} 
\end{aug}

\received{\smonth{11} \syear{2010}}
\revised{\smonth{9} \syear{2011}}

%
\begin{abstract}
The estimation of the total of an attribute defined over a continuous
planar domain is required in many applied settings, such as the
estimation of canopy coverage in the Monterano Nature Reserve in Italy.
If the design-based approach is considered, the scheme for the
placement of the sample sites over the domain is fundamental in order
to implement the survey. In real situations, a commonly adopted scheme
is based on partitioning the domain into suitable strata, in such a way
that a single sample site is uniformly placed (i.e., selected
with uniform probability density) in each stratum and sample sites are
independently located. Under mild conditions on the function
representing the target attribute, it is shown that this scheme gives
rise to an unbiased spatial total estimator which is ``superefficient''
with respect to the estimator based on the uniform placement of
independent sample sites over the domain. In addition, the large-sample
normality of the estimator is proven and variance estimation issues are
discussed.
\end{abstract}

%

\end{frontmatter}

\section{Introduction}\label{intro}

Applied scientists frequently deal with attributes defined on continuous
spatial domains. In this framework, if the design-based approach is assumed,
the target attribute may be expressed as a fixed bounded function $y$ taking
values on the study region $A$ (a suitable subset of the plane). In the
simplest case, $y(u)$ may represent the value of the attribute at $u\in A$.
As an example, in an environmental survey, $y(u)$ could be the air-borne
pollutant level at the sample site $u$ on a landscape. In a more structured
setting, $y(u)$ may also describe the ``attribute density'' arising
from the
selected spatial sampling design [this topic is extensively considered and
explained in Chapter 10 of \citet{GreVal08}]. As an
example, by
supposing a fixed-radius circular plot sampling in a forest survey, $y(u)$
could represent the number of trees lying in the plot centered at the sample
site $u$ [up to a known proportionality constant; see \citet{GreVal08}, pages
328--332]. In this case, under the design-based
approach, the
population universe is constituted by a \textit{continuum} (ideally by the
noncountable set of sample sites on $A$) and the inference is actually
carried out by assuming the so-called ``continuous-population'' paradigm.
This approach has been extensively considered in recent years on the
basis of the seminal papers by \citet{deGter90}, \citet{Cor93}
and \citet{BrudeG97}.

In the described framework, the estimation goal is usually focused on the
spatial total, that is,
%
%
\begin{equation}\label{11}
T=\int_Ay(u) \,\mathrm{d}u
\end{equation}
[see, e.g., \citet{Ste97} and Chapter 10 of \citet{GreVal08}].
Indeed, as emphasized by \citet{Ste97}, the integral representation in
(\ref{11})
embraces a general family of population parameters, such as means,
proportions or distribution functions. In order to estimate $T$, the key
problem of the design-based approach is the selection of an appropriate
sampling strategy. As usual, it is assumed that the sampling strategy
includes the joint selection of a suitable estimator and the corresponding
scheme for the placement of the $n$ sample sites on the study region $A$.
Since an integral representation for $T$ holds, it is quite evident
that the
estimation problem may be rephrased in terms of the Monte Carlo integration
theory. Interestingly, known Monte Carlo integration strategies are
equivalent to the sampling strategies which are commonly adopted in
environmental and ecological studies [Barabesi (\citeyear{Bar03}, \citeyear{Bar07}),
\citet{GreVal08}, page~327]. Similar Monte Carlo integration
approaches to
parameter estimation occur in very different research areas, such as in
stereology [see, e.g., the monograph by \citet{BadJen05}] or in
computer graphics [see, e.g., \citet{Agaetal03}].

\begin{figure}

\includegraphics{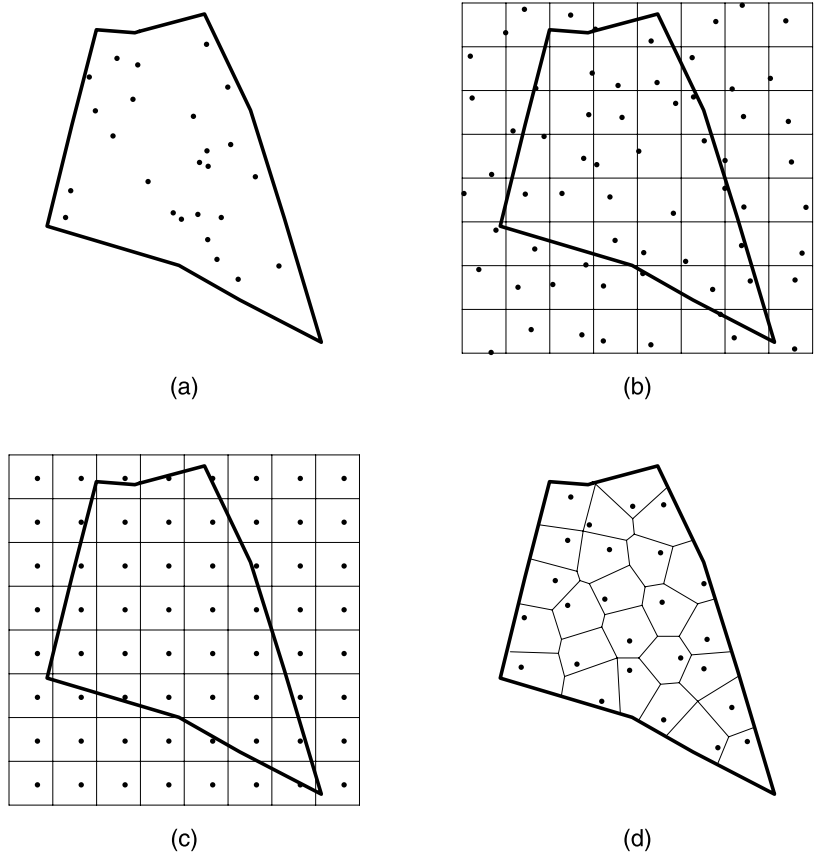}

\caption{Placement of $n=25$ sample sites over a study
region $A$ according to the URS scheme \textup{(a)}, the TSS scheme
\textup{(b)}, the SGS scheme \textup{(c)} and the SS scheme based on equal-size strata obtained by
means of the Brus, Sp{\"a}tjens and
de~Gruijter (\protect\citeyear{BruSpadeG99}) algorithm \textup{(d)}. For the TSS and
SGS schemes, the set $R$ is chosen as a rectangle and the tessellation
is based on squares, while $n$ represents the mean number of sample
sites in $A$.} \label{f1}
\end{figure}

The basic reference sampling scheme for selecting the sample sites is the
Uniform Random Sampling (URS), which constitutes the continuous-population
analog to simple random sampling from a finite population [\citet{Cor93}].
Under URS, the $n$ sample sites are independently and uniformly
selected on
$A$ [Figure~\ref{f1}(a)]. Despite its simplicity, URS may be not suitable
in practice
since it may produce an uneven coverage of the study region and the
corresponding unbiased estimator of $T$ displays\vspace*{1pt} a variance of order
$n^{-1}$, that is, the variance decreases to $0$ at the rate
$n^{-1}$ as
$n\rightarrow\infty$. In any case, URS is often considered a helpful
benchmark to compare the performance of more refined schemes.

%
%

In order to overcome the drawbacks involved with URS, a sampling scheme
frequently adopted in environmental studies is the so-called Tessellation
Stratified Sampling (TSS) [see, e.g., \citet{Ste97} and the
\citet{Env02}, page 63]. The TSS scheme is initially
implemented by
superimposing a suitable set $R$ onto the study region in such a way that
$A\subseteq R$ and by introducing the analytical extension $y_e$ of $y$ on
$R$. Formally, the analytical extension is defined as $y_e(u)=y(u)$ if
$u\in
A$ and $y_e(u)=0$ if $u\in R\setminus A$. Obviously, in this setting
(\ref{11}) may
be conveniently rewritten as $T=\int_Ry_e(u)\,\mathrm{d}u$. Subsequently, a regular
tessellation of~$R$ is carried out and one sample site is independently and
uniformly selected in each tessellation element [Figure \ref{f1}(b)]. The
theoretical
properties of the TSS scheme have been explained in detail
[Barabesi and Marcheselli (\citeyear{BarMar03}, \citeyear{BarMar05N1}, \citeyear{BarMar05N2}, \citeyear{BarMar08})]. The scheme gives rise to
an unbiased
estimator for~$T$ with variance of order $n^{-\gamma}$, where
$\gamma\in(1,2]$. Hence, the estimator under TSS is ``superefficient'' since
its variance decreases to $0$ faster than the variance of the estimator under
URS as $n\rightarrow\infty$. The parameter $\gamma$ depends on the
degree of
regularity of the analytical extension $y_e$: for instance, it turns
out that
$\gamma=2$ for smooth functions, but it can occur that $\gamma=3/2$
even for
noncontinuous and rather irregular functions [Barabesi and Marcheselli (\citeyear{BarMar03}, \citeyear{BarMar05N1}, \citeyear{BarMar08})]. Hence, the variance of the estimator for $T$
decreases to
$0$ faster as $y_e$ becomes more regular. In addition, consistent variance
estimation is available if $y_e$ is a differentiable function on $R$
[Barabesi and Marcheselli (\citeyear{BarMar03}, \citeyear{BarMar08})]. In order to avoid the difficulties
involved in the variance estimation, \citet{CorTho95} and \citet{Ste97} suggest modifying the TSS scheme by randomly shifting the
tessellation. The randomized TSS scheme does not involve extra-sampling
effort and allows for unbiased variance estimation without any
restriction on
the function $y_e$ [\citet{Ste97}], even if the consistency of the variance
estimator has not been proved. \citet{BarFra11} show that the
TSS scheme and its randomized modification produce estimators for $T$ with
identical variance convergence rates.

A further frequently-considered scheme is Systematic Grid Sampling\break (SGS),
which constitutes a systematic version of the TSS scheme [see,
e.g., \citet{ValAffGre09} and the \citet{Env02}, page 70]. The SGS scheme
provides the continuous-popula\-tion analog to systematic sampling from a
finite population as considered by \citet{MadMad44} [see also
\citet{DOr03}, as to systematic sampling of a spatial finite
population]. Similarly to
TSS, the SGS scheme requires a regular tessellation of $R$ and the extension
of the function $y$ on $R$. However, under SGS, a sample site is uniformly
generated in the reference tessellation element and it is systematically
repeated in the other tessellation elements [see Figure \ref{f1}(c)]. The
SGS scheme
is commonly adopted in stereology and gives rise to an unbiased
``superefficient'' estimator for~$T$, under certain smoothness
conditions on
$y_e$ [\citet{Cru93}, \citet{BadJen05}, page 159].
However, the
variance of the estimator tends to be extremely elevated when the periodicity
in the function~$y_e$ is ``in phase'' with the
tessellation elements [\citet{BadJen05}, Chapter 13].

Even if the TSS and SGS schemes allow for an even coverage of the study
region and give rise to unbiased ``superefficient'' estimation for $T$, these
schemes suffer due to two main technical drawbacks which are related to the
analytical extension of $y$ on $R$. Indeed, the sample sites are actually
placed on $R$ (not on~$A$) and, hence, the number of sample sites on
$A$ is a~random variable, unless $A$ is exactly tessellated. Obviously, the
tessellation may be selected in such a way that the mean number of sample
sites on~$A$ equals the prefixed sample size $n$ [as an example, this
procedure is adopted for Figure~\ref{f1}(b) and~(c)].
However, the task is generally
undesirable for field scientists, who usually require reproducible designs.
Moreover, even if the function $y$ is regular on $A$, the function
$y_e$ is
likely to be not continuous on the boundary of $A$ (and hence on $R$), unless
$y$ is null on this boundary. As previously explained, the lack of regularity
considerably reduces the efficiency of the estimators for $T$.

An alternative way to face the whole setting may be based on stratification
methods involving the ``one-per-stratum'' placement of the sample
sites. More
precisely, under the ``one-per-stratum'' Stratified Sampling (SS), the study
region $A$ is partitioned into $n$ suitable strata and one sample site is
independently and uniformly selected in each stratum [Figure \ref
{f1}(d)]. The scheme
constitutes the continuous-population counterpart to the classic
``one-per-stratum'' stratified sampling in the finite-population
setting [see,
e.g., \citet{Coc46}]. Moreover, the SS scheme generalizes the
TSS scheme when
$A$ coincides with $R$ and the tessellation is noncongruent. The scheme is
commonly adopted for environmental and agricultural surveys [see,
e.g., \citet{WalBrudeG10}
and the references therein].

In this paper, it is proven that the SS scheme produces an unbiased
``superefficient'' estimator for $T$, which shares the variance
properties of
the estimator under the TSS scheme. However, in contrast with the TSS
and SGS
schemes, when the SS scheme is adopted the $n$ sample sites are exactly
placed on $A$ and no analytical extension of $y$ is introduced.
Moreover, in
real surveys spatial stratification is often demanded in practice,
owing to
geographical or administrative convenience and tessellation-based methods
would not be applicable. In addition, there exist ad hoc
algorithms for
partitioning the study region into strata (eventually of the same size) with
suitable geometrical and statistical properties [\citet{BruSpadeG99},
\citet{WalBrudeG10}]. Finally, even if schemes with
more than a single sample site
per stratum may be considered, it is apparent that the benefits arising from
the full force of the stratification are achieved by adopting the
``one-per-stratum'' allocation. In any case, the ``two-per-stratum'' scheme
will be briefly considered since it produces unbiased and consistent variance
estimation.

Even if the achieved results may be applied to a broad range of different
data sets collected on a continuous spatial domain, the motivating practical
setting of the paper originates from an experiment dealing with canopy
coverage estimation in the Monterano Nature Reserve. Owing to the complex
boundary mosaic of this forest, the estimation approach based on forest
polygon delineation and area mensuration in the GIS environment may produce
omission and commission errors (which tend to be systematic) in the image
interpretation. In order to overcome these shortcomings, a survey procedure
based on line-intercept sampling which just involves the measurement of the
intersections of linear transects with forest patches is considered. As to
the Monterano Nature Reserve, forest researchers collected data by placing
transect midpoints according to the SS scheme with equal-size strata by means
of the \citet{BruSpadeG99}. Hence, the results of this paper
may be
suitably applied in order to provide point and interval estimation of canopy
coverage.

\section{Spatial total estimation}\label{sec2}
As pointed out in the \hyperref[intro]{Introduction}, the benchmark for comparing different
schemes is the URS and, hence, spatial total estimation under this
scheme is
briefly described. If $U_1,U_2,\ldots,U_n$ are $n$ i.i.d. random variables
representing the sample-site locations, in such a way that each $U_i$ is
uniformly distributed on $A$, the usual unbiased estimator for $T$
under URS
[see, e.g., \citet{Cor93}] is
%
%
\begin{equation}\label{21}
\widetilde{T}_n=\frac{a(A)}{n} \sum_{i=1}^ny(U_i),
\end{equation}
where $a(\cdot)$ denotes the area of a set in $\bbr^2$ (technically,
$a$ represents the Lebesgue measure in $\bbr^2$). The variance of
the estimator in (\ref{21}) is given by
\[
\operatorname{Var}[\widetilde{T}_n]=\frac{1}{n} \bigl(a(A)S-T^2\bigr),
\]
where $S=\int_Ay(u)^2\,\mathrm{d}u$, and, hence, it turns out that
$\operatorname{Var}[\widetilde{T}_n]=O(n^{-1})$. As usual, if $\{a_n\}$ and
$\{b_n\}$
represent two positive sequences, $a_n=O(b_n)$ means that the ratio $a_n/b_n$
is bounded for all $n$.

Under the SS scheme, the study region $A$ is partitioned in $n$ strata~$A_1,\allowbreak A_2,\ldots, A_n$, in such a way that each stratum is connected and
compact (in a topological sense), and one sample-site location is independently
and uniformly selected in each stratum. Therefore, let us suppose that the
sample-site locations are represented by the random variables
$V_1,V_2,\ldots,V_n$. According to the continuous Horvitz--Thompson Theorem
[\citet{Cor93}], an unbiased estimator for (\ref{11}) under SS is
%
%
\begin{equation}\label{22}
\widehat{T}_n=\sum_{i=1}^na(A_i)y(V_i)
\end{equation}
with variance
\[
\sigma_n^2=\operatorname{Var}[\widehat{T}_n]=\sum_{i=1}^na(A_i)^2
\operatorname{Var}[y(V_i)]=%
\sum_{i=1}^na(A_i)S_i-\sum_{i=1}^nT_i^2,
\]
where $T_i=\int_{A_i}y(u)\,\mathrm{d}u$ and $S_i=\int_{A_i}y(u)^2\,
\mathrm{d}u$.

In order to assess the variance properties of the estimator in
(\ref{22}), let us assume
that $y$ is a H\"older function on $A$, that is, a function
satisfying the
condition
\[
|y(u)-y(v)|\leq H\|u-v\|^\alpha,
\]
where $H<\infty$, $\alpha\in(0,1]$ and $u,v\in A$, while
\mbox{$\|\cdot\|$} denotes the usual norm in $\bbr^2$, that is,
$\|u-v\|$ denotes the distance between the points $u$ and $v$.
Obviously,\vadjust{\goodbreak} $y$~reduces to a Lipschitz function for the special case
$\alpha=1$. The family of H\"older functions is very large [for more
details, see, e.g., \citet{Eva10}, page 254]. Indeed, it is at
once apparent that H\"older functions are continuous. In addition, from
the above definition, it also follows that the family of H\"older
functions contains the family of Lipschitz functions, which in turn
contains the family of continuously differentiable functions.
Informally speaking, the H\"older condition quantifies the local
variation of the function $y$, in such a way that the index $\alpha$
may be interpreted as the corresponding ``degree of local continuity.''
Hence, the family of H\"older functions encompasses ``smooth''
functions, as well as functions displaying a very irregular behavior.
As a matter of fact, there exist H\"older functions which are
continuous, but nowhere differentiable.

By assuming that
\[
\operatorname{diam}(B)={\sup_{u,v\in B}}\|u-v\|
\]
represents the diameter of a given set $B$, that is, the
largest distance between
two points in $B$, let
\[
d_n=\max_{i=1,2,\ldots,n}\operatorname{diam}(A_i)
\]
be the maximum diameter of the $A_i$'s. Hence, let us consider the condition
%
%
\begin{equation}\label{23}
d_n^2\leq bn^{-1},
\end{equation}
where $b>0$ is a bounded constant. Since
\[
a(A_i)\leq\operatorname{diam}(A_i)^2\leq d^2_n,\qquad i=1,2,\ldots,n,
\]
condition (\ref{23}) implies that
\[
a(A_i)\leq bn^{-1},\qquad i=1,2,\ldots,n.
\]
In addition, let us also consider the condition
%
%
\begin{equation}\label{24}
a(A_i)\geq cn^{-1},\qquad i=1,2,\ldots,n,
\end{equation}
where $c>0$ is a bounded constant. It should be remarked that
condition~(\ref{23}) simply requires that the stratification be performed by
assuming quite ``homogeneous'' strata, that is, avoiding strata having
``stretched'' shapes and in such a way that no ``large'' strata are
admitted as $n\rightarrow \infty$. In addition, condition (\ref{24})
actually ensures that too ``small'' strata are in turn avoided as
$n\rightarrow\infty$. These requirements are likely to hold for
practical choices of $A_1,A_2,\ldots,A_n$. Obviously, condition
(\ref{24}) is always satisfied with equal-size strata, that is, when
$a(A_i)=a(A)/n$ for $i=1,2,\ldots,n$.

On the basis of Result \ref{Result1} in the \hyperref[app]{Appendix}, by assuming that $y$ is a H\"older
function and that condition (\ref{23}) holds, it turns out that
\[
\sigma_n^2=O(n^{-1-\alpha}).
\]
Hence, the SS scheme may lead to a noticeable estimation improvement with
respect to the basic URS scheme. The best variance order $n^{-2}$ is achieved
when $y$ is a Lipschitz\vadjust{\goodbreak} function. In any case, the SS scheme produces more
efficient estimation with respect to URS for each $\alpha$ value as
$n\rightarrow\infty$. The gain may be remarkable since in many real surveys
$\alpha$ is likely to be about one---for example, as to the canopy coverage
estimation considered in Section \ref{sec4}; see the discussion after the
formula in (\ref{41}).

The achieved variance properties may be extended to a larger class of
functions. More precisely, let $y$ be a piecewise H\"older function on
$A$, that is, there exists a finite partition of $A$ in such a way that~$y$
is a H\"older function on each partition element and the partition
boundary is rectifiable, that is, in practical terms the boundary is
``smooth.'' This setting is of real interest, since~$y$ often belongs
to this function family when~$y$ represents the ``attribute density''
as defined in the \hyperref[intro]{Introduction}. Thus, by assuming
that~$y$ is a piecewise H\"older function and that condition (\ref{23})
holds, on the basis of Result \ref{Result2} in the \hyperref[app]{Appendix}, it turns out that
\[
\sigma_n^2=O\bigl(n^{-\min(1+\alpha,3/2)}\bigr).
\]
Hence, even if the gain is lessened owing to the discontinuity of the
function~$y$, the performance of the SS scheme is in turn considerable. In
this case, the best variance order $n^{-3/2}$ is achieved when $y$ is a
piecewise H\"older function with $\alpha\geq1/2$. In turn, the SS
scheme is
preferable with respect to the URS scheme for each $\alpha$.

As to the large-sample normality of the estimator in (\ref
{22}), on the basis of Result~\ref{Result3}
in the \hyperref[app]{Appendix}, by assuming that $y$ is a H\"older function and that
conditions in (\ref{23}) and in (\ref{24}) hold, it
follows that
\[
\frac{\widehat{T}_n-T}{\sigma_n}\stackrel{\mathcal{L}}{\longrightarrow}N(0,1)
\]
as $n\rightarrow\infty$. This convergence result holds even if $y$ is a
piecewise H\"older function (see Remark \ref{Remark2} in the \hyperref[app]{Appendix}). These
findings on
the variance properties and the large-sample normality of the estimator
in (\ref{22}) are in complete
agreement with the results obtained by Barabesi and Marcheselli (\citeyear{BarMar03}, \citeyear{BarMar05N1}, \citeyear{BarMar08}) and \citet{BarFra11} under TSS. Indeed, the
TSS scheme
may be considered a special case of the SS scheme when $A$ coincides
with $R$
and the strata correspond to the elements of the
tessellation.\looseness=-1

It should be finally emphasized that for each $n$ the variance of the
estimator under URS is greater than or equal to the variance of the estimator
under SS when the strata are of the same size, that is, it
holds that
\[
\operatorname{Var}[\widetilde{T}_n]\geq\operatorname{Var}[\widehat{T}_n].
\]
Indeed, in this case the previous inequality is verified since
$T=\sum_{i=1}^nT_i$ and $S=\sum_{i=1}^nS_i$, while the inequality
\[
\sum_{i=1}^nT_i^2\geq\frac{1}{n} \Biggl(\sum_{i=1}^nT_i\Biggr)^2
\]
obviously holds.\vadjust{\goodbreak}

\section{Variance estimation}\label{sec3}
The estimation of $\sigma_n^2$ is not a trivial task, since a single
observation per stratum is available and the strata generally do not
constitute a regular tessellation, that is, the strata display
different sizes and shapes. In such a setting, estimators relying on
contrast-based techniques---such as the proposals by Barabesi and
Marcheselli (\citeyear{BarMar03}, \citeyear{BarMar08}) under TSS or the
proposal by \citet{SteOls03} under randomized TSS---seem quite
difficult to implement. However, a simple estimator may be obtained by
treating the sample as if it were obtained under the URS scheme. A
similar procedure is suggested by \citet{SteOls03} under the
randomized TSS scheme. Hence, a \textit{na\"\i ve} estimator for
$\sigma _n^2$ is given by
%
%
\begin{equation}\label{31}
\widehat{\sigma}_n^2=\frac{n}{n-1} \sum_{i=1}^n\biggl(a(A_i)y(V_i)-\frac
{%
\widehat{T}_n}{n}\biggr)^2.
\end{equation}
Since
\[
\mathrm{E}[\widehat{\sigma}_n^2]=\sigma_n^2+\mathrm{B}[\widehat{\sigma}_n^2],
\]
where
\[
\mathrm{B}[\widehat{\sigma}_n^2]=\frac{n}{n-1} \sum_{i=1}^n\biggl(T_i-%
\frac{T}{n}\biggr)^2
\]
(Result \ref{Result4} in the \hyperref[app]{Appendix}), the estimator in (\ref{31}) is
positively biased. Moreover, if
$y$ is a H\"older function and condition (\ref{23}) holds, it follows that
\[
\mathrm{B}[\widehat{\sigma}_n^2]=O(n^{-1})
\]
(Result \ref{Result4} in the \hyperref[app]{Appendix}). Hence, even if B$[\widehat{\sigma}_n^2]$ vanishes
for large $n$, it might be of a larger order than that of $\sigma_n^2$.
Moreover, if condition (\ref{24}) holds, on the basis of Remark \ref{Remark1} in
the \hyperref[app]{Appendix},
it promptly turns out that
\[
\frac{\mathrm{B}[\widehat{\sigma}_n^2]}{\sigma_n^2}=O(n).
\]
In any case, the behavior of this type of estimator seems quite stable in
practice as emphasized by \citet{SteOls03}, even if its use may lead
to a marked overestimation of $\sigma_n^2$.

\begin{figure}

\includegraphics{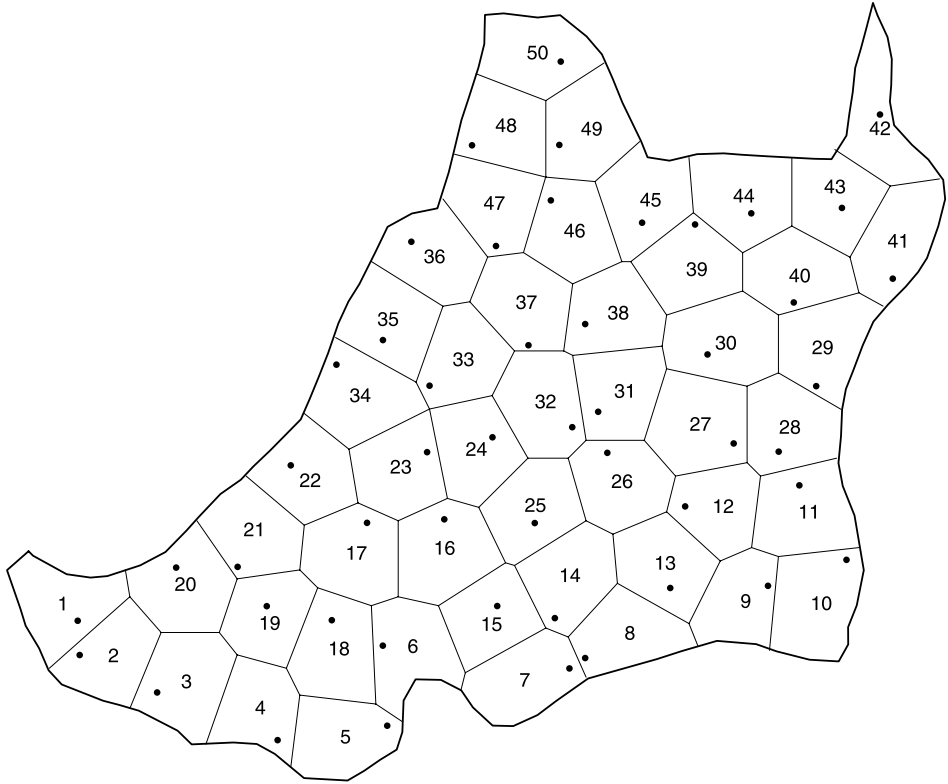}%
\vspace*{-3pt}
\caption{The Monterano Nature Reserve and the corresponding partition in
$n=50$ strata according to the SS scheme based on equal-size strata obtained
by means of the Brus, Sp{\"a}tjens and
de~Gruijter (\protect\citeyear{BruSpadeG99}) algorithm.}
\label{f2}
\vspace*{-3pt}
\end{figure}

For equal-size strata, an alternative estimator displaying more appealing
features may be proposed. The suggested estimator is given by
%
%
\begin{equation}\label{32}
\widetilde{\sigma}_n^2=\frac{a(A)^2}{2n^2} \Biggl(y(V_1)^2+%
\sum_{i=1}^{n-1}\bigl(y(V_i)-y(V_{i+1})\bigr)^2+y(V_n)^2\Biggr).
\end{equation}
Since
\[
\mathrm{E}[\widetilde{\sigma}_n^2]=\sigma_n^2+\mathrm{B}[\widetilde
{\sigma}_n^2],
\]
where
\[
\mathrm{B}[\widetilde{\sigma}_n^2]=\frac{1}{2} \Biggl(T_1^2+%
\sum_{i=1}^{n-1}(T_i-T_{i+1})^2+T_n^2\Biggr)
\]
(Result \ref{Result5} in the \hyperref[app]{Appendix}), the estimator in (\ref{32}) is
positively biased. By assuming
that
\[
D_n={\max_{i=1,2,\ldots,n-1}\sup_{u\in A_i,v\in A_{i+1}}}\|u-v\|,
\]
let us consider the condition
%
%
\begin{equation}\label{33}
D_n^2\leq kn^{-1}
\end{equation}
with $k>0$ a suitable bounded constant. Condition (\ref{33}) actually
requires that
the stratification be performed by indexing the strata in such a way that
$A_i$ and $A_{i+1}$ not be ``too far'' with respect to each other. In
practical situations, $A_i$ and $A_{i+1}$ may be generally chosen as
``neighbors,'' that is, in such a way that they share part of
their boundary. For
example, in the case study contained in Section \ref{sec4}, the partition
elements are
equally-sized strata which are indexed in such a way that the $i$th and the
$(i+1)$th strata have a~side in common (see Figure \ref{f2}). If $y$ is
a H\"older
function and conditions~(\ref{23}) and~(\ref{33}) are satisfied, it
follows that
\[
\mathrm{B}[\widetilde{\sigma}_n^2]=O(n^{-1-\alpha})
\]
and
\[
\limsup_n\biggl|\frac{\widetilde{\sigma}_n^2-\mathrm{E}[\widetilde{%
\sigma}_n^2]}{\sigma_n^2}\biggr|=0
\]
(Result \ref{Result5} in the \hyperref[app]{Appendix}). Hence, the bias order of the estimator in
(\ref{32}) is reduced
with respect to the estimator in (\ref{31}) and B$[\widetilde
{\sigma}_n^2]$ is of the same
order as that of $\sigma_n^2$. Moreover, if condition (\ref{24}) holds,
and on the
basis of Remark \ref{Remark1} in the \hyperref[app]{Appendix}, it follows that
\[
\frac{\mathrm{B}[\widetilde{\sigma}_n^2]}{\sigma_n^2}=O(n^{1-\alpha
}).
\]
Hence, when $y$ is a Lipschitz function, it turns out that
\[
0\leq\frac{\mathrm{B}[\widetilde{\sigma}_n^2]}{\sigma_n^2}\leq m,
\]
where $m>0$ is a suitable bounded constant, while
\[
1\leq\liminf_n\frac{\widetilde{\sigma}_n^2}{\sigma_n^2}\leq\limsup
_n\frac{%
\widetilde{\sigma}_n^2}{\sigma_n^2}\leq1+m
\]
(Result \ref{Result5} in the \hyperref[app]{Appendix}), that is, the estimator in
(\ref{32}) is large-sample
conservative.\vadjust{\goodbreak}

%
%

Finally, unbiased and consistent variance estimation is achieved if two
sample sites are placed in each stratum, that is, if the
``two-per-stratum'' SS
scheme is actually adopted. In this case, let us assume that $n$ is
even and
that a partition of the study region into $n/2$ strata is carried out.
Obviously, $n$ is now required to be even for comparison purposes with
respect to the ``one-per-stratum'' SS scheme. Moreover, let $V_{1,j}$ and
$V_{2,j}$ represent the two sample sites uniformly and independently selected
onto the $j$th stratum ($j=1,2,\ldots,n/2$). An unbiased estimator for $T$
is given by
%
%
\begin{equation}\label{34}
\widehat{T}_{2,n}=\frac{1}{2} \sum
_{j=1}^{n/2}a(A_j)\bigl(y(V_{1,j})+y(V_{2,j})\bigr),
\end{equation}
while its variance is
\[
\sigma_{2,n}^2=\operatorname{Var}[\widehat{T}_{2,n}]=\frac{1}{2} \sum
_{j=1}^{n/2}a(A_j)^2%
\operatorname{Var}[y(V_{1,j})]=\frac{\sigma_{n/2}^2}{2}.\vadjust{\goodbreak}
\]
Hence, an unbiased and consistent estimator for $\sigma_{2,n}^2$ is
%
%
\begin{equation}\label{35}
\widehat{\sigma}_{2,n}^2=\frac{1}{4} \sum_{j=1}^{n/2}a(A_j)^2\bigl(y(V_{1,%
j})-y(V_{2,j})\bigr)^2.
\end{equation}
Moreover, by considering Result \ref{Result1} in the \hyperref[app]{Appendix}, when condition (\ref
{23}) holds
it turns out that
\[
\sigma_{2,n}^2=O(n^{-1-\alpha})
\]
if $y$ is a H\"older function on $A$, while
\[
\sigma_{2,n}^2=O\bigl(n^{-\min(1+\alpha,3/2)}\bigr),
\]
if $y$ is a piecewise H\"older function on $A$. However, even if the
``two-per-stratum'' SS scheme provides in turn an unbiased ``superefficient''
estimator for $T$, it is at once apparent that an efficiency loss
occurs in
using a stratification based on $n/2$ strata rather than $n$ strata. In
addition, if the $n/2$ strata are split in such a way that each stratum is
partitioned into two substrata of equal sizes and $\widehat{T}_n$\vspace*{1pt} is computed
on the basis of this stratification, it is promptly shown that
$\operatorname{Var}[\widehat{T}_{2,n}]\geq\operatorname{Var}[\widehat{T}_n]$ on the
basis of the
discussion at the end of Section \ref{sec2}.

\section{An application to canopy coverage estimation}\label{sec4}
In order to illustrate an application of the SS scheme in an environmental
survey, the estimation of the canopy coverage in the Monterano Nature Reserve
has been considered. The Monterano Nature Reserve (which constitutes the
study region $A$ in this case) is located in the central part of Italy (Lazio
region) and its geographical boundary is depicted in Figure \ref{f2}.
The area of
the Monterano Nature Reserve is equal to $a(A)=1\mbox{,}045$~ha.

If $C\subset A$ represents the region inside $A$ covered by vegetation,
canopy coverage is simply defined as the area of $C$, that is,
in this case
$T=a(C)$. Canopy coverage constitutes a central indicator in forestry, as
emphasized by \citet{Bon89}. In order to estimate this quantity, replicated
line-intercept sampling is commonly adopted [\citet{Bar07}, \citet{BarMar08}]. More precisely, the replicated line-intercept sampling
protocol is carried out by selecting $n$ sample sites on $A$ and by
considering $n$ linear transects of fixed length $L$ with the same
orientation, in such a way that the transect midpoints are centered on each
sample site. According to \citet{BarMar08}, the canopy
coverage $T$ may be expressed as the integral of the ``attribute density,''
that is,
\[
T=a(C)=\frac{1}{L} \int_Al\bigl(C\cap t(u)\bigr) \,\mathrm{d}u,
\]
where $t(u)$ represents the set of points in a transect with midpoint
centered at the sample site $u$,\vadjust{\goodbreak} while $l(\cdot)$ denotes the length of
a set
in $\bbr$ (technically,~$l$ represents the Lebesgue measure in
$\bbr$). In this case, it follows that
\[
y(u)=\frac{1}{L} l\bigl(C\cap t(u)\bigr)
\]
is the length of the intersection of $t(u)$ with $C$, up to a known constant.
Hence, if the SS scheme with equal-size strata is adopted, the canopy
coverage estimator reduces to
%
%
\begin{equation}\label{41}
\widehat{T}_n=\frac{a(A)}{Ln} \sum_{i=1}^nl\bigl(C\cap t(V_i)\bigr),
\end{equation}
that is, the estimator in (\ref{41}) actually
represents the total sum of the intersection
lengths between $C$ and the $n$ transects, up to a known constant.
\citet{BarMar08} remark that if the boundary of $C$ is
rectifiable, $y$
is a H\"older function. Hence, if condition (\ref{23}) holds, it turns
out that
$\operatorname{Var}[\widehat{T}_n]=O(n^{-1-\alpha})$. In particular, if $C$
is given by the\vspace*{1pt}
union of circles or ellipses, it may be proven that
$\operatorname{Var}[\widehat{T}_n]=O(n^{-2+\varepsilon})$ where $\varepsilon>0$.
Moreover, if~$C$
is given by the union of polygons, it may be proven that
$\operatorname{Var}[\widehat{T}_n]=O(n^{-2})$. Hence, in real settings, the
estimator in (\ref{41}) may
be very efficient.

Even if the canopy coverage could be estimated by means of polygon
delineation on the basis of visual interpretation of remotely sensed imagery,
the procedure may typically produce errors and omissions [see,
e.g., \citet{CorChiTra04}]. So, in order to avoid the
interpretation drawbacks in the
estimation of forest features such as forest ecotone or canopy coverage,
\citet{CorChiTra04} suggest adopting replicated
line-intercept sampling.
Hence, for estimating canopy coverage in the Monterano Nature Reserve, the
replicated line-intercept sampling protocol has been implemented by assuming
the described procedure with $n=50$ transects with fixed direction and length
$L=200$ m [these choices are consistent with the study
by \citet{CorChiTra04}]. The transect midpoints (displayed in Figure \ref{f2})
have been placed by
adopting the SS scheme with equal-size strata obtained by using the
\citet{BruSpadeG99} algorithm. In this case, the estimator in (\ref
{41}) has given rise to the
estimate $660$ ha for the canopy coverage. Hence, $63.16$\% of the Monterano
Nature Reserve is covered by vegetation. In addition, variance
estimation has
been performed on the basis of (\ref{32}) by adopting a
sequential indexing of
strata with a common side (see Figure \ref{f2}). Accordingly, the
standard deviation
estimate is $58$ ha and a conservative confidence interval for canopy coverage
at the approximate $95$\% confidence level is $(647\mbox{ ha},674\mbox{ ha})$. Thus, a
conservative confidence interval at the same level for the percent coverage
is given by $(61.87\%,64.46\%)$.

\section{Concluding remarks}\label{sec5}

Under the design-based approach, the target attribute of many surveys
can be
conceptualized as a suitable fixed function $y$ defined on a given planar\vadjust{\goodbreak}
domain. This approach is usually described as the continuous-population
paradigm and it is especially suitable in environmental and ecological
frameworks [see, e.g., \citet{WilEri02} and
\citet{GreVal08}, page 2]. Indeed, in such spatial
contexts, it is not often possible to
achieve an area frame in order to apply the usual finite-population sampling
theory. Regrettably, practitioners frequently force the continuous-population
setting into the finite-population setting, owing to the lack of
results or
to the misunderstanding of the continuous-population paradigm.

The continuous-population paradigm requires implementation of an effective
probability sampling design to estimate the target parameter, usually the
total of the study attribute given by the integral of the function $y$.
Hence, a key decision is the choice of the sampling scheme for the placement
of sample sites. Schemes based on tessellation and stratification are widely
used in natural resource assessment and for environmental monitoring, since
evenly-spread sample sites over the study region often simplify
collection of
field data and estimation efficiency is usually increased. However, as
emphasized by \citet{WalBrudeG10}, schemes based on
tessellation methods
may involve several drawbacks and, hence, stratification schemes may
often be
preferable.

In the present paper it is shown that the ``one-per-stratum'' placement of
the sample sites produces an unbiased ``superefficient'' spatial total
estimator with respect to the uniform placement of independent sample sites.
Variance properties and convergence results for the suggested estimator are
given in a~purely design-based approach without assuming any super-population
model on the spatial correlation structure of the target attribute, as
usually considered for systematic and stratified sampling of a
two-dimensional population [see, e.g., \citet{Bel77} and
\citet{Bre}]. In
contrast, the present findings are achieved by assuming very mild conditions
on the function $y$ (which are likely to be met in any real survey) and by
requiring simple conditions which avoid strata of too small or too large
sizes, as well as strata with stretched shapes.

\begin{appendix}\label{app}
\section*{Appendix}

\begin{Result}\label{Result1}
Let $y$ be a H\"older function. Hence, since each $A_i$ is assumed
to be connected and $y$ is a continuous function, there exists $\bar
{u}_i\in
A_i$ for each $i=1,2,\ldots,n$ such that
\[
y(\bar{u}_i)=\mathrm{E}[y(V_i)].
\]
Accordingly, since the H\"older condition holds for $y$, we obtain
\begin{eqnarray*}
\operatorname{Var}[y(V_i)]&=&\mathrm{E}\bigl[\bigl(y(V_i)-\mathrm{E}[y(V_i)]\bigr)^2\bigr]
=%
\mathrm{E}\bigl[\bigl(y(V_i)-y(\bar{u}_i)\bigr)^2\bigr]\\
&\leq& H^2\mathrm{E}[\|V_i-\bar{u}_i\|^{2\alpha}]\\
&\leq& H^2\operatorname{diam}(A_i)^{2\alpha}\leq H^2d^{2\alpha}_n.
\end{eqnarray*}
Hence, it holds that
\[
\sigma_n^2=\sum_{i=1}^na(A_i)^2\operatorname{Var}[y(V_i)]\leq
H^2d_n^{2\alpha} \sum_{i=1}^na(A_i)^2.
\]
Since $a(A_i)\leq d^2_n$ and $\sum_{i=1}^na(A_i)=a(A)$, it also turns
out that
\[
\sigma_n^2\leq H^2d_n^{2+2\alpha}a(A).
\]
In addition, if condition (\ref{23}) holds, it follows that
\[
\sigma_n^2\leq b^{1+\alpha}H^2a(A)n^{-1-\alpha},
\]
that is, $\sigma_n^2=O(n^{-1-\alpha})$.
\end{Result}
\begin{Result}\label{Result2}
Let $y$ be a piecewise H\"older function on $A$. Moreover, by
denoting $B$ as the boundary of the partition, let us assume that
\[
I=\{i\dvtx A_i\cap B\neq\varnothing\}.
\]
Moreover, if
\[
M=\sup_{u\in A}|y(u)|,
\]
then $\operatorname{Var}[y(V_i)]\leq M^2$ and it holds that
\[
\sum_{i\in I}a(A_i)^2\operatorname{Var}[y(V_i)]\leq M^2d_n^4
\operatorname{card}(I),
\]
where card$(\cdot)$ denotes cardinality of a set. Since $B$ is rectifiable,
it turns out that
\[
\operatorname{card}(I)\leq C_1n^{1/2},
\]
where $C_1>0$ is a suitable bounded constant. Hence, by assuming
condition~(\ref{23}), it follows that
\[
\sum_{i\in I}a(A_i)^2\operatorname{Var}[y(V_i)]\leq b^2M^2C_1n^{-3/2}.
\]
In addition, since $y$ is a H\"older function on $A_i$ for $i\notin I$, by
assuming the achievements in Result \ref{Result1}, it holds that
\[
\sum_{i\notin I}a(A_i)^2\operatorname{Var}[y(V_i)]\leq
H^2d_n^{4+2\alpha}\operatorname{card}(I^c).
\]
Moreover, it turns out that
\[
\operatorname{card}(I^c)\leq C_2n,
\]
where $C_2>0$ is a suitable bounded constant. Thus, by assuming condition~(\ref{23}), it follows that
\[
\sum_{i\notin I}a(A_i)^2\operatorname{Var}[y(V_i)]\leq
b^{2+\alpha}H^2C_2n^{-1-\alpha}.
\]
Hence, it is finally seen that
\[
\sigma_n^2\leq b^2M^2C_1n^{-3/2}+b^{2+\alpha}H^2C_2n^{-1-\alpha},
\]
that is, $\sigma_n^2=O(n^{-\min(1+\alpha,3/2)})$.
\end{Result}
\begin{Remark}\label{Remark1}
If $y$ is not a constant function on $A$ and if
condition~(\ref{23}) holds, it follows that
\[
\liminf_n\sum_{i=1}^n\operatorname{Var}[y(V_i)]\geq M_y,
\]
where $M_y>0$ is a bounded constant depending on $y$. Hence, from condition~(\ref{24}) we have
\[
\sigma_n^2\geq c^2M_yn^{-2}.
\]
\end{Remark}
\begin{Result}\label{Result3}
Let $y$ be a H\"older function and let us assume that conditions~(\ref{23})
and~(\ref{24}) hold. In order to prove the large-sample
normality of the estimator in~(\ref{22}), it suffices to verify the
Lyapunov condition, that is,
\[
\lim_{n\rightarrow\infty}\frac{v_n}{\sigma_n^3}=0,
\]
where
\[
v_n=\sum_{i=1}^na(A_i)^3\mathrm{E}\bigl[|y(V_i)-\mathrm{E}[y(V_i)]|^3\bigr].
\]
By assuming the notation and the findings of Result \ref{Result1}, it turns out that
\[
v_n\leq
d_n^2 \sum_{i=1}^na(A_i)^2\mathrm{E}\bigl[|y(V_i)-y(\bar{u}_i)|^3\bigr].
\]
Moreover, since the H\"older condition holds for $y$, it also follows that
\begin{eqnarray*}
v_n&\leq&
Hd_n^2 \sum_{i=1}^na(A_i)^2 \mathrm{E}\bigl[\bigl(y(V_i)-y(\bar{u}_i)\bigr)^2 \|V_i-
\bar{u}_i\|^\alpha\bigr]\\
&\leq& Hd^{2+\alpha}_n
\sum_{i=1}^na(A_i)^2\operatorname{Var}[y(V_i)]=Hd_n^{2+\alpha}\sigma_n^2
\end{eqnarray*}
and, hence,
\[
\frac{v_n}{\sigma_n^3}\leq\frac{Hd_n^{2+\alpha}}{\sigma_n}.
\]
Thus, on the basis of condition (\ref{23}) and Remark \ref{Remark1}, it follows that
\[
\frac{v_n}{\sigma_n^3}\leq c^{-1}b^{1+\alpha/2}HM_y^{-1/2}n^{-\alpha/2}
\]
and, hence, the Lyapunov condition is proven.
\end{Result}
\begin{Remark}\label{Remark2}
The large-sample normality of the estimator in~(\ref{22}) may be proven even if
$y$ is a piecewise H\"older function and conditions~(\ref{23}) and~(\ref
{24}) hold. This
result may be shown in a general setting by verifying the Raikov condition.
\end{Remark}
\begin{Result}\label{Result4}
Since the estimator in (\ref{31}) may be
rewritten as
\[
\widehat{\sigma}_n^2=\frac{n}{n-1} \sum_{i=1}^na(A_i)^2y(V_i)^2-%
\frac{1}{n-1} \widehat{T}_n^2,
\]
it follows that
\begin{eqnarray*}
\mathrm{E}[\widehat{\sigma}_n^2]&=&\frac{n}{n-1} \sum_{i=1}^na(A_i)^2%
\mathrm{E}[y(V_i)^2]-\frac{1}{n-1}\mathrm{E}[\widehat{T}_n^2]\\
&=&\frac{n}{n-1} \Biggl(\sigma_n^2+\sum_{i=1}^nT_i^2\Biggr)-\frac{1}{n-1} (%
\sigma_n^2+T^2)=\sigma_n^2+\mathrm{B}[\widehat{\sigma}_n^2].
\end{eqnarray*}
On the basis of the notation and the findings of Result \ref{Result1}, since
\[
T_i\leq Ma(A_i),
\]
where $M$ is defined in Result \ref{Result2}, it turns out that
\begin{eqnarray*}
\sum_{i=1}^n\biggl(T_i-\frac{T}{n}\biggr)^2&\leq&
2 \sum_{i=1}^nT_i^2+2T^2n^{-1}\\
&\leq& 2M^2 \sum_{i=1}^na(A_i)^2+2T^2n^{-1}\\
&\leq& 2M^2a(A)d_n^2 +2T^2n^{-1}.
\end{eqnarray*}
Hence, if condition (\ref{23}) holds, it follows that
\[
\sum_{i=1}^n\biggl(T_i-\frac{T}{n}\biggr)^2\leq2\bigl(bM^2a(A)+T^2\bigr)n^{-1}
\]
and, hence, B$[\widehat{\sigma}_n^2]=O(n^{-1})$.
\end{Result}
\begin{Result}\label{Result5}
Since the estimator in (\ref{32}) may be
rewritten as
\[
\widetilde{\sigma}_n^2=\frac{a(A)^2}{n^2} \Biggl(\sum_{i=1}^ny(V_i)^2-%
\sum_{i=1}^{n-1}y(V_i)y(V_{i+1})\Biggr),
\]
it follows that
\begin{eqnarray*}
\mathrm{E}[\widetilde{\sigma}_n^2]&=&\frac{a(A)^2}{n^2} \Biggl(\sum
_{i=1}^n%
\mathrm{E}[y(V_i)^2]-\sum_{i=1}^{n-1}\mathrm{E}[y(V_i)]%
\mathrm{E}[y(V_{i+1})]\Biggr)\\
&=&\sigma_n^2+\sum_{i=1}^nT_i^2-\sum_{i=1}^{n-1}T_iT_{i+1}=\sigma_n^2+%
\mathrm{B}[\widetilde{\sigma}_n^2].
\end{eqnarray*}
Moreover, if $y$ is a H\"older function, we have
\begin{eqnarray*}
(T_i-T_{i+1})^2&\leq&\frac{a(A)^2}{n^2} \bigl(y(\bar{u}_i)-y(\bar
{u}_{i+1})\bigr)^2\\
&\leq&\frac{a(A)^2}{n^2} H^2\|\bar{u}_i-\bar{u}_{i+1}\|^{2\alpha}\\
&\leq& H^2a(A)^2D_n^{2\alpha}n^{-2}.
\end{eqnarray*}
Hence, if condition (\ref{33}) holds, it turns out that
\[
(T_i-T_{i+1})^2\leq k^\alpha H^2a(A)^2n^{-2-\alpha}.
\]
Thus, it follows that
\[
\mathrm{B}[\widetilde{\sigma}_n^2]\leq M^2a(A)^2n^{-2}+k^\alpha
H^2a(A)^2n^{-1-\alpha},
\]
where $M$ is defined in Result \ref{Result2}, and, hence,
B$[\widetilde{\sigma}_n^2]=O(n^{-1-\alpha})$. Moreover, if condition
(\ref{24})
holds, owing to Remark \ref{Remark1}, we obtain
\begin{eqnarray*}
\sum_{n \geq
1}P\biggl(\biggl|\frac{\widetilde{\sigma}_n^2-\mathrm{E}[\widetilde{%
\sigma}_n^2]}{\sigma_n^2}\biggr|>\varepsilon\biggr)&\leq&\sum_{n \geq
1}\frac{1}{\sigma_n^8\varepsilon^4}\mathrm{E}\bigl[(\widetilde{\sigma}_n^2-%
\mathrm{E}[\widetilde{\sigma}_n^2])^4\bigr]\\
&\leq&\sum_{n \geq
1}\frac{n^4}{c^4M_y^8\varepsilon^4}\mathrm{E}\bigl[(\widetilde{\sigma}_n^2-%
\mathrm{E}[\widetilde{\sigma}_n^2])^4\bigr]\\
&\leq&\frac{C_3}{c^4M_y^8\varepsilon^4} \sum_{n \geq1}\frac
{1}{n^2}<\infty
\end{eqnarray*}
since
\[
\mathrm{E}\bigl[(\widetilde{\sigma}_n^2-\mathrm{E}[\widetilde{\sigma
}_n^2])^4\bigr]%
\leq C_3n^{-6},
\]
where $C_3>0$ is a suitable bounded constant. Hence, it follows that
\[
\lim_{n\rightarrow\infty}\frac{\widetilde{\sigma}_n^2-\mathrm{E}[%
\widetilde{\sigma}_n^2]}{\sigma_n^2}=0.
\]
\end{Result}
\end{appendix}

\section*{Acknowledgments}

The authors would like to thank Professor Lorenzo Fattorini for many
helpful suggestions and Professor Luca Pratelli for useful advice in
the proofs of the \hyperref[app]{Appendix} results. They are also grateful to Professor
Piermaria Corona for providing the Monterano Nature Reserve data set.


%

\printaddresses

\end{document}